\documentclass[mathleft, fleqn, 
]{an}
\usepackage{graphicx}
\usepackage{times}
\usepackage{amsmath}
\usepackage{natbib}
\bibpunct{(}{)}{;}{a}{}{,}
\def\fd{\hbox{$.\!\!^{\rm d}$}}

\def\degr{\hbox{$^\circ$}}
\def\sun{\hbox{$\odot$}}
\def\kms{\mbox{km\,s$^{-1}$\/}}
\def\bfcyg{\mbox{BF~Cyg\,}}
\def\zand{\mbox{Z~And\,}}
\def\agpeg{\mbox{AG~Peg\,}}
\def\pcyg{\mbox{P~Cyg\,}}

\begin{document}

\Pagespan{789}{}
\Yearpublication{2006}%
\Yearsubmission{2005}%
\Month{11}%
\Volume{999}%
\Issue{88}%

\title{
Transient accretion disc-like envelope in the symbiotic binary BF Cygni \\
during its 2006 -- 2015 optical outburst}

\author{N.\,A.\,Tomov\inst{1}\fnmsep\thanks{Corresponding author:
  \email{tomov@astro.bas.bg}}
\and  M.\,T.\,Tomova\inst{1}
\and  D.\,V.\,Bisikalo\inst{2}
}
\titlerunning{Accretion envelope in BF Cygni}
\authorrunning{N.\,A.\,Tomov et al.}
\institute{
Institute of Astronomy and National Astronomical Observatory,\\ Bulgarian Academy of Sciences,  POBox 136, 4700 Smolyan, Bulgaria
\and 
Institute of Astronomy of the Russian Academy of Sciences,\\ 48 Pyatnitskaya Str., 119017 Moscow, Russia
}

\received{2015 Mar 31}
\accepted{2015 Jul 17}
\publonline{XXXX}


\abstract{
The optical light of the symbiotic binary \bfcyg during its last eruption after 2006 shows orbital variations because of an eclipse of the outbursting compact object. The first orbital minimum is deeper than the following ones. Moreover, the Balmer profiles of this system acquired additional satellite components indicating bipolar collimated outflow at one time between the first and second orbital minima. This behaviour is interpreted in the framework of the model of collimated stellar wind from the outbursting object. It is supposed that one extended disc-like envelope covering the accretion disc of the compact object and collimating its stellar wind forms in the period between the first and second minima. The uneclipsed part of this envelope is responsible for the decrease of the depth of the orbital minimum. The calculated $UBVR_{\mathrm C}I_{\mathrm C}$ fluxes of this uneclipsed part are in agreement with the observed residual of the depths of the first and second orbital minima. The parameters of the envelope require that it is the main emitting region of the line H$\alpha$ but the H$\alpha$ profile is less determined from its rotation and mostly from other mechanisms. It is concluded that the envelope is a transient nebular region and its destruction determines the increase of the depth of the orbital minimum with fading of the optical light.}

\maketitle


	\section{Introduction}

Symbiotic stars are interacting binaries consisting of a normal cool giant or Mira and a hot compact object accreting matter from the atmosphere of the cool companion. As a result of accretion the compact object undergoes recurrent optical outbursts. The photometric variability of the symbiotic stars is determined from both the orbital motion and the outburst events of the accretor which are often accompanied by intensive mass ejection in the form of optically thick shells, stellar wind and bipolar collimated jets. Some of these binaries are eclipsing systems whose eclipses are observed in both of their states -- the quiescent and active ones.

The symbiotic system \bfcyg is an eclipsing binary consisting of a late-type cool component classified as an M5\,III giant \citep{KFC87,MS}, a hot luminous compact object with a temperature of about 10$^5$ K and an extended surrounding nebula \citep{Sk05}. Its orbital period is 757\fd2 based on both photometric \citep{Mik87} and radial velocity \citep{Fekel+01} data.

The historical light curve of \bfcyg shows three types of activity: a very prolonged outburst, lasting for decades and similar to that of the symbiotic novae, several eruptions such as those observed in the classical symbiotic stars and sudden rapid brightenings lasting for a small part of the orbital period \citep{Sk+97}.
\citet{LF06} suggested that some magnetic cycle in the outer layers of the giant component is connected to the origin of the outbursts. During the previous eruption in 1989 -- 1993 an expanding shell with absorption lines mostly of \mbox{Fe\,{\sc ii}} was observed \citep{Cassatella+92}. The eclipses in this system are good seen during its previous outburst, its recent outburst begun in 2006 and the quiescent period between them \citep{Sk+15}.


\begin{figure*}
\centering
	\includegraphics[angle=-90,width=0.95\textwidth]{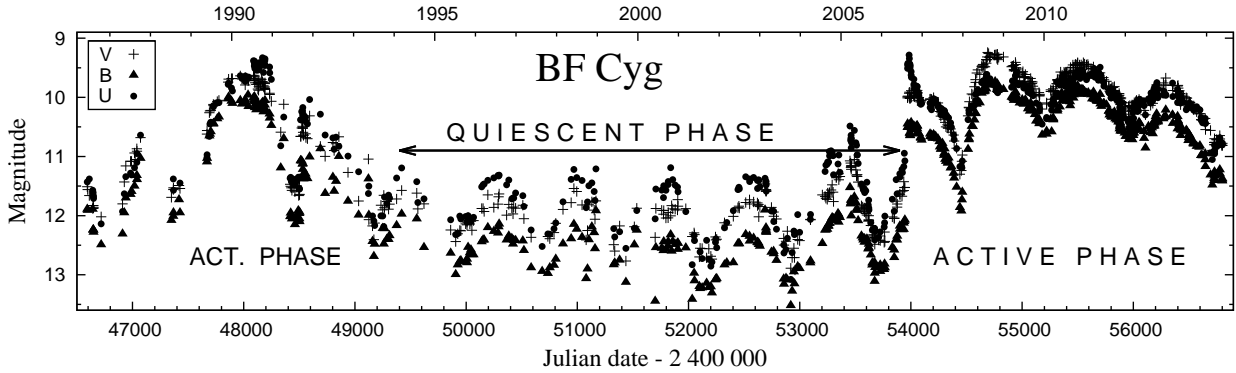}
 \caption{The $UBV$ light curves of \bfcyg during its outbursts after 1989 and 2006 and the quiescent stage between them. Adopted from \citet{Sk+15}.}
 \label{bf_lc}
\end{figure*}

The recent 2006 -- 2015 outburst of \bfcyg was first reported by \citet{Munari+06}.
The line spectrum was investigated by \citet{McKeever+11}. Up to now the brightness has had four orbital minima during this outburst determined from eclipses \citep{Sk+15}. The first of them is deeper than the following ones which means that the geometrical structure of the eclipsed outbursting compact object has changed for the time between the first and second orbital minima. Moreover, at one time between the first and second minima, since June 2009, bipolar collimated ejection appeared which lasted till April 2013. Balmer lines H$\alpha$ and H$\beta$ acquired satellite components with a velocity of about 400~\kms\, \citep{Sk+13,Sk+15}. The blueshifted H$\beta$ component initially appeared in absorption and after September 2011 went into emission. Satellite components indicating a collimated ejection were thus observed for the first time in this system. Moreover, the H$\beta$ profile contained two additional variable \pcyg absorptions with lower velocities of $-30$ to $-90$~\kms\, and $-250$~\kms\, \citep{TTB14}. All of these data were interpreted in the framework of the model of collimated stellar wind by the outbursting object \citep{TTB14}. 


In this work we analyse the optical light curves of \bfcyg during its 2006 -- 2015 outburst with the aim to propose one quantitative interpretation of the decrease of the depth of the orbital minimum (eclipse) in the framework of the same model suggesting thus a possible answer to the question why did satellite components appear in the spectrum just in between the first and second orbital minima. An other task is to interprete all features of the H$\alpha$ profile in the light of this idea.

	\section{The behaviour of the optical light}

The eclipse of the compact object in \bfcyg occurs at the time of the inferior conjunction of the giant according to the ephemeris of \citet{Fekel+01} 
\[ {\rm JD(Min)} = 2\,451\,395\fd2 + 757\fd2 \times {\rm E}\,. \]  
The system shows periodic wave-like orbital variation in its quiescent state similar to those of \agpeg \citep{Sk+12}. The analysis of the spectral energy distribution by \citet{Sk05} shows that in quiescence the $UBV$ continuum fluxes of the circumbinary nebula are comparable to the continuum fluxes of the hot compact object. Then the decrease of the brightness at the phase of the orbital minimum is determined from both an eclipse of the compact object and an occultation of a part of the circumbinary nebula. At other orbital phases the nebula is  partially occulted too.

Eclipses are observed in the \bfcyg system during its outbursts as well \citep{Sk+12}. Its last outburst began in July -- August 2006 and continues up to the present time (Fig.~\ref{bf_lc}) when several orbital minima were detected caused by eclipses \citep{Sk+15}. 
The first minimum is deeper than the following ones. After the first maximum the light decreases in different way compared to the next maxima. Initially the $U$ light falls steepy at about one magnitude. In our view it is most probably due to decrease of the mass-loss rate of the outbursting component which moves the level of the observed photosphere back to the star 
and redistributes the continuum emission from longer wavelengths towards the UV region. After the initial steepy fall the light begins to decrease more slowly and the basic reason for this decrease is the occultation of the circumbinary nebula.
Assuming a  decrease of the mass-loss rate, a part of the ejected material can remain within the gravitational potential of the compact object and  begin to accrete \citep{TTB14}. Observational indication of the increase of the mass-loss rate can be an appearance of absorption component of the spectral lines. During the rise of the light to the second maximum the mass-loss rate was probably increased as proposed by the spectral data of \citet{Siviero+12}. Strong \pcyg components of some Balmer and metal lines appeared after 2008 February 10. The light reached its second maximum at the end of 2008. Some time after that the Balmer lines acquired satellite components indicating bipolar collimated outflow \citep{Sk+13}.

\section{Interpretation}


\begin{table*}[!htb]
\caption{Dereddened fluxes of \bfcyg in units of $10^{-12}$ erg\,cm$^{-2}$\,s$^{-1}$\,\AA$^{-1}$ with their inner uncertainties at the times of the orbital minima under consideration.}
\label{tab:1}
\centering
\begin{tabular}{lllllll}
\hline
    \noalign{\smallskip}
Date & JD$-$2400000 & $F_U$ & $F_B$ & $F_V$ & $F_{R_{\mathrm C}}$ & $F_{I_{\mathrm C}}$ \\
    \noalign{\smallskip}
\hline
    \noalign{\smallskip}
Dec. 16, 2007 & 54451.220 & 0.720$\pm$0.037 & 0.533$\pm$0.015 & 0.367$\pm$0.011 & 0.330$\pm$0.009 & 0.291$\pm$0.009 \\
Dec. 13, 2009 & 55179.202 & 1.375$\pm$0.065 & 1.522$\pm$0.044 & 0.883$\pm$0.025 & 0.613$\pm$0.017 & 0.511$\pm$0.015 \\
\hline
\end{tabular}
\end{table*}

A scenario to interpret the spectrum of \zand during its 2000 -- 2013 active phase involving two stages of the evolution of the outbursting compact object was presented in the works of \citet{TTB12,TTB14}. It is supposed that there is geometrically thin disc from wind accretion surrounding the compact object in the system in its quiescence and during the first outburst of the active phase. At the end of the first outburst some part of the ejected material accretes again and because of conservation of its initial angular momentum falls into the disc. Thus one extended disc-like envelope covering the disc forms, which locates at a greater distance from the orbital plane than the accretion disc itself. In such a way the second stage of the evolution of the outbursting object begins, which is related to the following outbursts of the active phase. During the following outbursts the envelope can collimate the outflowing material and bipolar collimated outflow (collimated stellar wind) from the outbursting object can form.

According to our model the decrease of the depth of the orbital minimum implies an increase of the emission of the uneclipsed geometrical structure of the compact object. We suppose that a thin accretion disc from wind accretion has initially existed in the system \bfcyg and an extended disc-like envelope has formed after that as a result of decrease of the mass-loss rate and accretion of material from the potential well of the compact object in the period between the two orbital minima. We suppose that this envelope has collimated the stellar wind later. Some part of it is not eclipsed during the second and following minima determining their smaller depth.
The jet is approximated with a spherical sector. The continuum flux determined by recombinations and free-free transitions of a spherical sector in a wind with a constant velocity $\upsilon$ is 
$F_{\lambda}^{\rm cont} \sim 1/\upsilon^2$. Then the emission of the jets is negligible because of the high velocity of the outflow. 

Our aim is to calculate the $UBVR_{\mathrm C}I_{\mathrm C}$ emission of such a model structure and to compare it with the observed residual of the depths of the first and second minima. If the model emission (the uneclipsed part of the envelope) is close to the observed residual, we can conclude that the formation of a disc-like envelope collimating the stellar wind of the compact object is a possible reason for both the appearance of the satellite components of the spectral lines and decrease of the depth of the orbital minimum of \bfcyg. After the third minimum the system's brightness fades and the depth of the orbital minimum increases again. We suppose that former is probably due to decrease of the optical flux of the outbursting object and destruction of the envelope and the latter -- to destruction of the envelope only.

\section{Calculation}

Gas dynamical modeling shows that a disc from wind accretion with a typical radius of 50 -- 60~R$_{\sun}$ and a mass of $5\times 10^{-7}$~M$_{\sun}$ exists in one binary system with parameters close to those of \zand in its quiescent state \citep{Bis+02,T+10,T+11}. During the outburst the wind of the compact object ``strips'' the disc and ejects some part of its mass. At the end of the outburst some part of the ejected mass locates in the potential well of the compact object. After the cessation of the wind it begins to accrete again creating an extended disc-like envelope with a mass smaller than the mass of the initial accretion disc.

The system \bfcyg has parameters very close to those of \zand --  orbital period, masses of the components and mass-loss rate of the cool giant \citep{Fekel+01}. Then we suppose that a disc from accretion of a stellar wind with a size and a mass close to that in the system \zand exists in \bfcyg in its quiescent state. During the outburst the newly appeared disc-like envelope should have smaller mass. We accept an inner radius of the envelope of 25~R$_{\sun}$, equal to the radius of the pseudophotosphere of the outbursting compact object on 2008 October 23 according to \citet{Sk+15}, an outer radius of $\sim\!\!150$~R$_{\sun}$ very close to the mean size of the Roche lobe of the compact object, a height of the envelope of $\sim\!\!170$~R$_{\sun}$ and a mass of $\sim\!\!3.5\times 10^{-7}$~M$_{\sun}$. Its mean density is thus $7.5\times 10^{10}$~cm$^{-3}$. 
In reality the shape of this envelope differs from cylinder. With use of the orbital period and masses of the components of \bfcyg of \citet{Fekel+01} for the binary separation we obtain 492~R$_{\sun}$. \citet{Fekel+01} propose a radius of the cool giant of $70\pm32$~R$_{\sun}$ at a distance to the system of 2~kpc, and \citet{Sk05} $\sim\!\!150$~R$_{\sun}$ at a distance of 3.8~kpc. \citet{Fekel+01} propose an orbit inclination of 75\degr, and \citet{Sk+97} $70\div90\degr$. We accept an inclination of 75\degr. In this case the radius of the giant should be large to provide eclipse and we accept a radius of 150~R$_{\sun}$ according to \citet{Sk05}. With this radius about a half of the volume of the disc-like envelope will be visible during the eclipse at orbital phase 0.0.

To calculate the residual of the depths of the first and second orbital minima we used photoelectric and CCD data from the work of \citet{Sk+12}. The data of 2007 December 16 at the time of the first minimum are an arithmetical mean of several estimates taken in December 2007. At the time of the second minimum we used the $UBV$ data of 2009 December 13 and $R_{\mathrm C}I_{\mathrm C}$ data of 2010 January 23 since there was no infrared data in December. The stellar magnitudes were converted into continuum fluxes using the calibration from the book of \citet{Mih}. The $U$ flux was not corrected for the energy distribution of \bfcyg in the region of the Balmer jump since we were not provided with spectral data in this region. The $UBV$ fluxes were not corrected for the emission lines for the same reason -- absence of spectral data in the $UBV$ region. All fluxes were corrected for the interstellar reddening $E(B-V) = 0.35$ \citep{Sk05} with use of the approach of \citet{Cardelli+89} and are listed in Table~\ref{tab:1}.



\begin{table*}
\caption{$UBVR_{\mathrm C}I_{\mathrm C}$ continuum fluxes and emission measure of the different regions in the circumbinary nebula. The fluxes are in units of $10^{-12}$ erg\,cm$^{-2}$\,s$^{-1}$\,\AA$^{-1}$ and emission measure in $10^{61}$ cm$^{-3}$.}
\label{tab:2}
\begin{center}
\begin{tabular}{lllllll}
\hline
    \noalign{\smallskip}
Emitting region & $F_U$ & $F_B$ & $F_V$ & $F_{R_{\mathrm C}}$ & $F_{I_{\mathrm C}}$ & $n_{\rm e}^2 V$ \\
    \noalign{\smallskip}
\hline
    \noalign{\smallskip}
Whole nebula & 2.184 & 0.913 & 0.768 & 0.647 & 0.491 & 2.60$^a$ \\
\\Disc-like envelope & 1.856 & 0.776 & 0.653 & 0.550 & 0.417 & 2.21 \\
\\Uneclipsed part & 0.928 & 0.388 & 0.327 & 0.275 & 0.208 & 1.10 \\
\\Residual of the depths & 0.655$\pm$0.075 & 0.989$\pm$0.046 & 0.516$\pm$0.027 & 0.283$\pm$0.019 & 0.220$\pm$0.017 & \\
\\$r^b$ & 42 & $-$61 & $-$37 & $-$3 & $-$5 & \\
\hline
    \noalign{\smallskip}
\end{tabular}
\end{center}

$^a$ The data are for 2008 October 23 in the work of Skopal et al. (2015).\\
$^b$ $r=(\mathrm U - \mathrm R)/\mathrm R$ in per cent; U -- Uneclipsed part, R -- Residual of the depths.

\end{table*}

We will calculate the emission of the uneclipsed part of the disc-like envelope at orbital phase 0.0 and will compare it with the observed residual of the depths of the first and second orbital minima in the photometric bands $UBVR_{\mathrm C}I_{\mathrm C}$. Our unpublished high resolution data taken at that time do not show presence of the line \mbox{He\,{\sc ii}} 4686 in the spectrum of \bfcyg. Then we will assume that helium in the nebula of \bfcyg is ionized and the nebular continuum is emitted by hydrogen and neutral helium. The continuum flux determined by recombinations and free-free transitions is given by

\begin{eqnarray}
	F_{\lambda}&=&\frac{1+a({\rm {He}})}{4\pi d^2}
	\left[ \gamma_\nu({\rm H^o},T_{\rm e})\, +
	 a({\rm {He}}) \gamma_\nu({\rm He^o},T_{\rm e})\,\right] n_{\rm e}^2 V
	 \nonumber  \\
 & &	\times	\frac{c}{\lambda^2} 10^{-8}\,,
\label{formula1}
\end{eqnarray}

\noindent where $a({\rm {He}})$ is helium abundance relative to hydrogen, $d$ is a distance to the system, $\gamma_\nu({\rm H^o},T_{\rm e})$ and $\gamma_\nu({\rm He^o},T_{\rm e})$ are continuum emission coefficients of hydrogen and helium, $n_{\rm e}$ electron density and $V$ volume of the emitting region.

We accept an electron temperature in emitting region $T_{\rm e} = 30\,000$~K equal to the mean temperature in the nebula according to \citet{Sk+15}. In such a case we used continuum emission coefficients at the positions of the $UBVR_{\mathrm C}I_{\mathrm C}$ photometric bands for this temperature from the paper of \citet{Ferland80} and the book of \citet{Pottasch84}. We took the arithmetical mean of the values of the hydrogen coefficient on both sides of the Balmer limit at the position of the $U$ band and accepted helium abundance of 0.1 \citep{VN94}.
The fluxes and emission measure of the whole circumbinary nebula, our model envelope and its uneclipsed part at orbital phase 0.0 are presented in Table~\ref{tab:2}. The emission measure of the nebula was taken from \citet{Sk+15} and its fluxes were calculated by us. The residual of the depths of the first and second orbital minima, obtained as residual of the fluxes in Table~\ref{tab:1}, is presented at the last row in Table~\ref{tab:2}. The model fluxes of the uneclipsed part are compared with this residual. It is seen that all fluxes of the uneclipsed part excepting only that in $B$ band are close to the residual. The reason for the greater difference in the $B$ band is rather unclear. The quiescent emission measure of \bfcyg at phase very close to the orbital photometric maximum is $3.1\times 10^{60}$~cm$^{-3}$ \citep{Sk05} and on 2008 October 23 it is $2.6\times 10^{61}$~cm$^{-3}$ \citep{Sk+15}. Its increase is thus $2.3\times 10^{61}$~cm$^{-3}$. Then it appears that the emission measure of the envelope of $2.2\times 10^{61}$~cm$^{-3}$ is almost equal to this increase.

\section{Discussion}

The evolution of the H$\alpha$ profile of \bfcyg during its outburst after 2006 was considered in the works of \citet{Sk+13} and \citet{Sk+15}. The observational data were taken in the period July 2006 -- April 2013.
A part of these data is shown in the work of \citep{TTB14} too. We interpreted the decrease of the depth of the orbital minimum with an appearance of a disc-like envelope whose emission measure is the greatest part of the emission measure of the whole circumbinary nebula. In such a case the Balmer lines should be emitted mainly by this envelope. The Keplerian velocity at its inner radius is 68~\kms\, and at the outer radius 27~\kms, using a mass of 0.6~M$_{\sun}$ for the compact object, according to \citet{Fekel+01}. Taking into consideration the orbit inclination, the radial velocities resulted from the rotation of the envelope should be 66 -- 26~\kms. If the H$\alpha$ profile was determined only by this rotation, it should be double-peaked with a peak separation of 52~\kms\, and a full width at zero intensity (FWZI) of 132~\kms. The observed peaks separation of the line during the time of the collimated outflow, however, ranged from 100~\kms\, to 150~\kms, which means that it was broadened more than only from the rotation of the envelope. The inner uncertainty of the measurement is not more than 5~\kms. A possible additional mechanism of line broadening is a gas turbulence caused by collision of the stellar wind of the outbursting component with the envelope.

The structure of the central emission of the H$\alpha$ line, however, was double-peaked before the first orbital minimum too \citep{Sk+13,Sk+15}, and the velocity of the dip between the two peaks was always of about 100~\kms. A \pcyg absorption with a velocity of about $-30$~\kms\, was present in the H$\beta$ line from June 2009 till May 2010 (see Fig~12 in the work of \citet{TTB14}). The velocity of this absorption was $-90$~\kms\, in September 2011 and after that it went into an absorption dip in the emission profile with a velocity of about $-30$~\kms\, again. It was concluded that it indicates a stellar wind which exists together with the collimated ejection \citep{TTB14}. We suppose that this wind produces the absorption dip in the emission profile of H$\alpha$ with a velocity of about $-100$~\kms.

\section{Conclusion}

The main results of our analysis can be summarised as follows:
\begin{itemize}

\item
We conclude that the decrease of the depth of the orbital photometric minimum of the eclipsing binary \bfcyg during its last optical outburst after 2006 and the appearance of satelliite components of the Balmer lines, indicating bipolar collimated outflow at one time between the first and second orbital minima, is probably due to formation of an extended disc-like envelope covering the accretion disc of the compact object and collimating its stellar wind. An increase of the uneclipsed part of the envelope is responsible for the decrease of the depth of the orbital minimum. We assume that the envelope results from diminution of the mass-loss rate of the compact object after the first light maximum and accretion of material from its potential well in the time between the first and second orbital minima. 

\item
We suppose that the shape of the envelope is close to torus with an inner radius of 25~R$_{\sun}$, equal to the radius of the pseudophotosphere of the outbursting object, an outer radius of $\sim\!\!150$~R$_{\sun}$ very close to the mean size of its Roche lobe and a height of $\sim\!\!170$~R$_{\sun}$. The mass of the envelope is accepted of $\sim\!\!3.5\times 10^{-7}$~M$_{\sun}$, its mean density thus amounts to $7.5\times 10^{10}$~cm$^{-3}$ and the emission measure to $2.21\times 10^{61}$~cm$^{-3}$. This emission measure is almost equal to the increase of the emission measure during the outburst.

In accordance with these parameters the central emission of the H$\alpha$ line should be produced mainly by this envelope but the analysis of its profile shows that it is additionaly broadened by a gas turbulence due to collision of the stellar wind with the envelope and its  double-peaked structure is determined mostly from an optically thick outflow producing an absorption dip in the emission profile.

\item
With use of a binary separation of 492~R$_{\sun}$ and inclination of the orbit of 75\degr\, \citep{Fekel+01} and a radius of the cool giant in the system of 150~R$_{\sun}$ \citep{Sk05} about a half of the volume of the disc-like envelope will be visible during the eclipse at orbital phase 0.0. We calculated the emitted by this volume $UBVR_{\mathrm C}I_{\mathrm C}$ fluxes and compared them with the observed residual of the depths of the first and second orbital minima. It turned out that the model fluxes are in agreement with the observed ones.

\item
We conclude that the increase of the depth of the orbital minimum with fading of the light during the outburst, on its side, is probably due to destruction of the disc-like envelope, which thus indicates its transient nature. 

\end{itemize}

\acknowledgements
The authors are grateful to the anonymous referee whose suggestions led to improve the paper.
This work has been supported by the Basic Research Program of the Presidium of the Russian Academy of Sciences, Russian Foundation for Basic Research 
(projects 14-02-00215, 14-29-06059), and the Russian and Bulgarian Academies of Sciences through a collaborative program in Basic Space Research.

\end{document}